\documentclass[showpacs,twocolumn,epsfig,pre]{revtex4-1}
\usepackage{graphicx}
\usepackage{amssymb}
\usepackage{color,soul}
\usepackage{amsmath}
\usepackage{latexsym}
\usepackage{amsmath}
\usepackage{latexsym}
\usepackage{array}
\usepackage{float}
\usepackage{amsfonts}
\usepackage{mathrsfs}
\usepackage{verbatim}
\usepackage{color}
\usepackage{eucal}

\newcommand{\ket}[1]{|#1\rangle}
\newcommand{\bra}[1]{\langle#1|}

\newcommand{\tr}{\mathrm{Tr}}

\begin{document}

\draft
\title{Information Fluctuation Theorem for an Open Quantum Bipartite System}
%FTs and thermodynamic inequalities for open quantum bipartite systems
\author{Jung Jun Park$^{1,2,3}$}
\email{hyok07@gmail.com}
\author{Hyunchul Nha$^{2}$}
\email{hyunchul.nha@qatar.tamu.edu}
\author{Sang Wook Kim$^{4}$}
\author{Vlatko Vedral$^{1,5}$}
\affiliation{$^1$Centre for Quantum Technologies, National University of Singapore, 3 Science Drive 2, 117543, Singapore}
\affiliation{$^2$Department of Physics, Texas A$\&$M University at Qatar, POBox 23874, Doha, Qatar}
\affiliation{$^3$Korea Institute for Advanced Study, Seoul 02455, Korea}
\affiliation{$^4$Department of Physics, Kyung Hee University, Seoul 02447, Republic of Korea}
\affiliation{$^5$Atomic and Laser Physics, Clarendon Laboratory, University of Oxford, Parks Road, Oxford OX13PU, United Kingdom}

\date{\today}

\begin{abstract}
\label{*abstract*}
We study an arbitrary non-equilibrium dynamics of a quantum bipartite system coupled to a reservoir. For its characterization, we present a fluctuation theorem (FT) that explicitly addresses the quantum correlation of subsystems during the thermodynamic evolution. To our aim, we designate the local and the global states altogether in the time-forward and the time-reversed transition probabilities. In view of the two-point measurement scheme, only the global states are subject to measurements whereas the local states are used only as an augmented information on the composite system. We specifically derive a FT in such a form that relates the entropy production of local systems in the time-forward transition to the change of quantum correlation in the time-reversed transition. This also leads to a useful thermodynamic inequality and we illustrate its advantage by an example of an isothermal process on Werner states. %Our illustration demonstrates that our characterization can be stronger than the existing ones. % and also discuss our results in view of Landauer's principle.
%For instance, the fundamental bounds of heat transferred from a heat reservoir to a quantum bipartite system involving the quantum mutual information are derived for both dissipative and non-dissipative processes.
 \end{abstract}
% 05.70.Ln Nonequilibrium and irreversible thermodynamics
% 03.67.-a Quantum information
% 05.30.-d Quantum statistical mechanics
% 89.70.Cf Entropy and other measures of information
% 03.65.Ta Foundations of quantum mechanics; measurement theory
% 05.70.-a Thermodynamics

\maketitle

\section{Introduction}
Given a physical system interacting with an environment, it is of fundamental and practical interest to know what constraints its thermodynamic evolution shall be subject to. 
One crucial tool to address such a question is the fluctuation theorem (FT) that characterizes the statistics of thermodynamic quantities such as entropy production and extractable work in an equality form valid even for non-equilibrium dynamics.
Generalizations of fluctuation relations to different physical scenarios have recently attracted great interest \cite{Evans1993, Gallavotti1995, Jarzynski1997, Crooks1999, Tasaki2000, Kurchan2000, Seifert2005, Talkner:2007aa, Crooks:2008aa, Esposito:2009aa, Campisi2009, Campisi2010, Campisi2011, Vedral:2012aa, Aberg:2018aa, Alhambra2016, Allahverdyan2014, Talkner2016, Dahlsten:2017aa, Quan2008, Gong2015, Chetrite:2012aa, Crooks2008, Kwon:2019aa, Holmes2018, Lostaglio:2015ab, Morris:2018aa, Crooks2008, Manzano2015, Sagawa2012,Sagawa2010,Funo2013,Jevtic2015,Manzano:2018aa,Morikuni:2011aa,Sagawa:2013aa,Park:2018aa,Ponmurugan:2010aa}. 
In particular, with the aim of extending our understanding to the quantum regime, a great deal of effort was made towards the identification of the role played by quantum principles in the emerging thermodynamic behaviors \cite{Goold:2016aa}.  
FTs were obtained to characterize the work statistics and the entropy production for quantum thermodynamical processes \cite{Quan2008,Manzano2015,Crooks:2008aa,Dahlsten:2017aa, Albash:2013aa}. Landauer's principle was rigorously formulated from the perspective of quantum statistical mechanics \cite{Reeb:2014aa}.
Other investigations were also made to consider a full quantum dynamics under the framework of quantum channels \cite{Aberg:2018aa,Alhambra2016,Kwon:2019aa, Holmes2018, Rastegin:2014aa}.
It was also discussed that the operational meaning of work, a basic notion of thermodynamics, becomes subtle in the quantum domain due to the quantum nature of dynamics (noncommutativity)  \cite{Allahverdyan2014,Talkner2016}.

On the other hand, FTs were also established to describe bipartite systems in both classical and quantum domains \cite{Sagawa2010, Morikuni:2011aa, Sagawa2012, Funo2013, Jevtic2015, Manzano:2018aa}. For instance, the work fluctuation via feedback control under information exchange between a system and a memory was addressed in \cite{Sagawa2010, Sagawa2012, Morikuni:2011aa}. Other FTs include the characterizations of heat fluctuation with information exchange between two subsystems \cite{Jevtic2015} and entropy production with information exchange between two subsystems \cite{Sagawa2012, Funo2013, Manzano:2018aa}.
In the quantum domain, various approaches were made to elucidate the thermodynamic aspects of quantum correlated systems. In \cite{Rio:2011aa,Sapienza:2019aa}, the quantum correlation was viewed as a thermodynamic resource in a single shot regime. It was also shown how to extract work by using quantum correlation \cite{Manzano:2018ab,Vitagliano:2018aa,Francica:2017aa,Perarnau-Llobet:2015aa,Friis:2016aa,Funo:2013aa,Park2013}. The work cost to create quantum correlation was also investigated in \cite{Bruschi:2015aa,Huber:2015aa}. 
%\st{ to demonstrate the thermodynamic laws} \cite{Bera:2017aa,Bera2018,Reeb:2014aa} \st{and the arrow of time }\cite{Jennings:2010aa,Partovi:2008aa}, \st{{\it etc}}.

%{\color{blue} (Alternatively, we may instead replace it by) }
%\textcolor{blue}{To elucidate the thermodynamic aspects of quantum correlated systems, various approaches were put forward to view correlations as a thermodynamic resource in a single shot regime \cite{Rio:2011aa,Sapienza:2019aa}, to show work extraction \cite{Manzano:2018ab,Vitagliano:2018aa,Francica:2017aa,Perarnau-Llobet:2015aa,Friis:2016aa,Funo:2013aa,Park2013} or work cost \cite{Bruschi:2015aa,Huber:2015aa} by using quantum correlation, and to demonstrate the thermodynamic laws \cite{Bera:2017aa,Bera2018,Reeb:2014aa} in terms of the thermodynamic arrow of time \cite{Jennings:2010aa,Partovi:2008aa}, {\it etc}.} %However, a general result of crucial importance in non-equilibrium thermodynamics, the so-called fluctuation theorem \cite{Deffner:2011aa}, was not yet obtained to incorporate the role of quantum correlation, which is the main purpose of this paper.

%\cite{Parrondo:2015aa, Goold:2016aa, Esposito:2010aa, Koski:2014aa,Chapman:2015aa,Deffner:2013aa,Horowitz:2015aa,Horowitz:2010aa,Strasberg:2017aa,Strasberg:2014aa,Horowitz:2014aa,Seifert:2012aa,Hartich:2014aa,Strasberg:2013aa,Abreu:2012aa,Ptaszynski:2019aa,Horowitz:2014ab,Esposito:2010aa,Barato:2014aa,Iyoda:2017aa}

%\cite{Sagawa2012,Sagawa2010,Funo2013,Jevtic2015,Manzano:2018aa,Morikuni:2011aa,Sagawa:2013aa,Kwon:2017aa,Park:2018aa,Ponmurugan:2010aa}

In the existing FTs for quantum bipartite systems, however, the role of quantum correlation present in an initial state was not explicitly investigated. Some works assumed that the initial state starts in a product state or in a classically correlated state \cite{Sagawa2010, Morikuni:2011aa, Sagawa2012, Funo2013, Jevtic2015, Manzano:2018aa, Ponmurugan:2010aa} while the quantum correlation may develop during the subsequent evolution. Moreover, the statistics of the information-thermodynamic quantities in the FT are typically obtained by performing a measurement on the subsystems at the initial and the final times under the standard two-point measurement protocol (TMP) \cite{Tasaki2000, Kurchan2000, Esposito:2009aa, Campisi2011, Vedral:2012aa}. 
The resulting information FTs with the TMP do not exhibit the quantum features of initial correlation because the measurement destroys quantum correlations. 
On the other hand, Deffner and Lutz obtained a FT based on the eigen-state measurement of the whole system before and after the evolution \cite{Deffner:2011aa}. This FT can be rephrased to manifest quantum correlation as well as local entropy production by dividing the whole system entropy into its constituents.

In this paper, we derive a FT for a non-equilibrium dynamics of a quantum bipartite system coupled to a reservoir, which explicitly addresses the quantum correlation during its evolution. 
To our end, we consider the statistics of the information-thermodynamic quantities in view of the TMP by measuring only the global state while we incorporate together the local states of the subsystems but without measuring them. 
This hybrid approach makes it possible to retain the quantum correlation of the subsystems intact in our formulation of FT \cite{Park:2017aa}. 
By using our approach explicitly considering quantum correlation, we first rephrase the FT by Deffner and Lutz \cite{Deffner:2011aa} but to a generalized form including the effect of absolute irreversibility \cite{Funo:2015aa,Murashita:2017aa}. 
Then, we specifically obtain a FT that relates the statistics of local entropy production in the time-forward transition to the change of quantum correlation in the time-reversed transition. 
This FT also leads to a thermodynamic inequality that can be useful to address the evolution of quantum bipartite systems. We illustrate the usefulness of our approach by studying an example, i.e. the isothermal process of a Werner state, a mixed quantum entangled state, to extract work by a time-dependent Hamiltonian. Our example clearly demonstrates that our thermodynamic inequality can be tighter than the existing ones.

\section{Preliminaries}

We start by considering a non-equilibrium process for a bipartite system $\rho_{AB} \in \mathcal{H}_A \otimes \mathcal{H}_B $ composed of two subsystems $A$ and $B$, with their local states $\rho_A \in \mathcal{H}_A$ and $\rho_B \in \mathcal{H}_B$, and a reservoir $\rho_R \in \mathcal{H}_R$. An arbitrary quantum bipartite state $\rho_{AB}^{\rm i}$ initially decoupled from $\rho_R^{\rm i}$ evolves into a final state $\rho_{ABR}^{\rm f} = U \rho_{AB}^{\rm i} \otimes \rho_R^{\rm i} U^{\dagger}$, where the process is described by a unitary operator $U$. We assume the initial state of $R$ to be a thermal equilibrium state $\rho_R^{\rm i} = e^{-\beta H_R}/Z_{\beta}$ with an inverse temperature $\beta$ and the corresponding partition function $Z_{\beta}$, where $H_R = \sum_r E^R_r \ket{r}\bra{r}$ is the Hamiltonian of $R$. The density operators of subsystems are given by the partial trace as $\rho_{A} = \tr_{B}(\rho_{AB})$ and $\rho_{B} = \tr_{A}(\rho_{AB})$, respectively. The spectral decompositions of $AB$, $A$, $B$, $R$ at the initial time $t_{{\rm i}}$ are denoted as $\rho_{AB}^{{\rm i}}  = \sum_{m} p_{m} \ket{m} \bra{m}$, $\rho_{A}^{{\rm i}}  = \sum_{a} p_{a} \ket{a} \bra{a}$, $\rho_{B}^{{\rm i}}= \sum_b p_{b}\ket{b} \bra{b}$, and $\rho_R^{\rm i}= \sum_{r} p_{r}\ket{r} \bra{r}$, respectively. 
Similarly, the final states are represented by using primed indices as 
$\rho_{AB}^{\rm f}  = \sum_{m^{\prime}} p_{m^{\prime}} \ket{m^{\prime}} \bra{m^{\prime}}$, $\rho_{A}^{\rm f}  = \sum_{a^{\prime}} p_{a^{\prime}} \ket{a^{\prime}} \bra{a^{\prime}}$, $\rho_{B}^{\rm f}= \sum_{b^{\prime}} p_{b^{\prime}}\ket{b^{\prime}} \bra{b^{\prime}}$, and $\rho_R^{\rm f}= \sum_{r^{\prime}} p_{r^{\prime}}\ket{r^{\prime}} \bra{r^{\prime}}$, respectively.

\subsection{Classical correlation}

Given a joint probability of two random variables, the classical stochastic mutual information $J$  can be defined as  $J(k,l) := \ln [p_{k,l}/p_{k}p_{l}]$, where $p_k$ and $p_l$ are the marginal probabilities of the joint probability $p_{k,l}$. We define the joint probability for subsystems $A$ and $B$ as $p_{a,b} := \bra{b}\bra{a}\rho_{AB}^{\rm i}\ket{b}\ket{a}$ at $t_{\rm i}$ and 
$p_{a',b'} := \bra{b'}\bra{a'}\rho_{AB}^{\rm f}\ket{b'}\ket{a'}$ at $t_{\rm f}$, respectively. 
The unaveraged classical mutual informations are then given by $J_{\rm i} = \ln [p_{a,b}/p_{a}p_{b}]$ at $t_{\rm i}$ and $J_{\rm f} = \ln [p_{a',b'}/p_{a'}p_{b'}]$ at $t_{\rm f}$ \cite{Sagawa2012}. The average of $J_{\rm i}$ is equal to the classical mutual information $\sum_{a,b} p_{a,b} J_{\rm i} = H_{\rm i}(A) + H_{\rm i}(B) - H_{\rm i}(A,B)$, where $H_{\rm i}(A)=-\sum_{a}p_{a}\log p_{a}$, $H_{\rm i}(B)=-\sum_{b}p_{b}\log p_{b}$, and $H_{\rm i}(A,B) = -\sum_{a,b}p_{a,b}\log p_{a,b}$ are the Shannon entropies of the local systems $A$, $B$, and the joint system $AB$, respectively. Similarly, the mutual information at the final time is given by using primed indices.  The classical mutual information represents only the classical correlation between the two subsystems, although the total system is set to be a quantum correlated state. Thus, the fluctuation theorem for quantum bipartite systems requires an information-thermodynamic quantity that describes the total correlation between the subsystems in an arbitrary quantum bipartite system.

\subsection{Quantum correlation}
In order to resolve this issue, we introduce a mutual information content $I$. Instead of the joint probability $p_{k,l}$, the probability $p_{m}$ of the total system is exploited to define $I_{\rm i}$ as $I_{m,a,b} :=  {\rm ln} [p_{m}/p_{a}p_{b}]$ at $t_{\rm i}$, and similarly at $t_{\rm f}$ by using primed indices. Comparing it to the classical stochastic mutual information $J$, it holds one more index $m$ representing the global state of the total system. To find the average mutual information content, the joint probability that involves three indices $m,a,b$ is required. Since the projectors of the total system $\Pi_{m} := \ket{m}\bra{m}$ in general do not commute with the products of the projectors of the subsystems $\Pi_{a} := \ket{a}\bra{a}$ and $\Pi_{b} := \ket{b}\bra{b}$, we are not allowed to define a joint probability that the total system is found in the state $\ket{m}$ and each subsystem is found in the state $\ket{a}$ for $A$ and $\ket{b}$ for $B$ at the same time \cite{Dirac1945}. Thus, using the conditional probability \cite{Margenau:1961aa} that the subsystems are in $\ket{a,b} := \ket{a} \otimes \ket{b}$ provided that the total system is in $\ket{m}$, a joint probability may be defined as
\begin{equation}
p_{m,a,b} :=\bra{m}\rho_{AB}^{\rm i}\ket{m}|\langle m | a, b \rangle|^2, \label{jp}
\end{equation}
where $p_{m} = \bra{m}\rho_{AB}^{\rm i}\ket{m}$ and $p_{a,b|m} = |\langle m | a,b \rangle|^2$ \cite{Kirkwood1933,Dirac1945,Barut1957,Margenau:1961aa,Allahverdyan2014}. 
The marginal probabilities of the subsystems can be well given by $p_{a} = \sum_{m,b} p_{m,a,b}$ and $p_{b} = \sum_{m,a} p_{m,a,b}$, respectively. The joint probability yields the average of $I_{\rm i}$ as $\langle I_{\rm i} \rangle = \sum_{m,a,b} p_{m,a,b} I_{\rm i} = S(\rho_A^{\rm i}) + S(\rho_B^{\rm i}) - S(\rho_{AB}^{\rm i})$, where $S(\rho) = -\tr(\rho \ln \rho)$ is the von Neumann entropy. $\langle I_{\rm i} \rangle$ then equals the quantum mutual information $I(A:B)$ of $\rho_{AB}^{\rm i}$ which is a measure of the total correlation of a quantum bipartite system \cite{Nielsen2000}. Similarly, we can define the joint probability and the mutual information at $t_{\rm f}$ by using primed indices and the final state $\rho_{AB}^{\rm f}$.

\subsection{Time-forward transition probability}
As we have earlier mentioned, the joint probability evaluated by the conventional TMP has its drawbacks \cite{Allahverdyan2014,Talkner2016}. 
In particular, the measurement performed at the initial time will destroy quantum correlation unless a nonlocal measurement is employed. This prevents us from characterizing the role of initial correlation in the subsequent thermodynamic behaviors. To include all the informational-thermodynamic quantities appropriately, we introduce a joint probability by adopting the TMP based on the measurements of only the global states while including together the probability distributions of the subsystems without measurement. 
The probability for the system $AB$ and the reservoir $R$ to be found in $\ket{m}$ and $\ket{r}$ at $t_{\rm i}$ and $\ket{m'}$ and $\ket{r'}$ at $t_{\rm f}$ is given by 
\begin{equation}
p_{m,m';r,r'} = |\bra{m',r'} U \ket{m,r}|^2 p_m p_r.\label{jp2}
\end{equation}
We then multiply the conditional probabilities $p_{a,b|m}$ and $p_{a',b'|m'}$ by $p_{m,m';r,r'}$ to define the probability that the subsystem $A$ and $B$ are found in the states $\ket{a}$ ($\ket{a'}$) and $\ket{b}$ ($\ket{b'}$) provided that the system $AB$ and the reservoir $R$ are in the states $\ket{m}$ ($\ket{m'}$) and $\ket{r}$ ($\ket{r'}$) at $t_{\rm i}$ ($t_{\rm f}$). That is, 
\begin{equation}
p_{m,a,b,m',a',b';r,r'} = p_{m,m';r,r'}|\langle m | a, b \rangle|^2 |\langle m' | a', b' \rangle|^2. \label{jp3}
\end{equation}
 We again emphasize that only the global state of the total system is measured in the TMP without measuring the subsystems directly. To see the validity of the probability defined this way, we sum $p_{m,a,b,m',a',b';r,r'}$ over all $m',a',b',r'$. We then find $\sum_{m',a',b',r'} p_{m,a,b,m',a',b';r,r'} = p_{m,a,b;r} = |\langle m | a, b \rangle|^2 p_mp_r$. In the same way, we can verify other marginal probabilities $p_{m',a',b';r'}$, $p_m$, {\it etc}. As in all other time-local approaches, we can also consider the multiple point measurement with infinitesimal time steps. Our definition can be straightforwardly extended to $P_{m_0,m_1,...,r_0,r_1,...}|\langle m_0 | a_0, b_0 \rangle|^2 |\langle m_1 | a_1, b_1 \rangle|^2 ...$ %\textbf{To figure out the new entities in detail, see the example in the following paragraph below Eq. \eqref{ineq2}.}

\subsection{Time-reversed transition probability}
As a crucial ingredient in FTs, we discuss a joint probability in a time-reversed transition. Corresponding to $U$ and $\rho_{AB}^{\rm f}$, suppose that a unitary operator $\tilde{U}$ describes a time-reversed process of an initial state $\tilde{\rho}_{AB}^{\rm i} = \sum_{\tilde{m}'} \tilde{p}_{m'} \ket{\tilde{m}'}\bra{\tilde{m}'}$, where $\ket{\tilde{m}'} = \Theta \ket{m'}$. $\Theta$ is a time-reversal operator. The joint probability of the time-reversed process is then given by
\begin{equation}
\tilde{p}_{m',a',b',m,a,b;r',r} =\tilde{p}_{m',m;r',r}  |\langle m' | a', b' \rangle|^2 |\langle m | a, b \rangle|^2\label{jp4},
\end{equation}
where $\tilde{p}_{m',m;r',r} = |\bra{m,r} U^{\dagger} \ket{m',r'}|^2 \tilde{p}_{m'}\tilde{p}_{r'}$ \cite{Esposito:2009aa}. It will be shown later that this probability is linked to the time-forward joint probability by a detailed fluctuation theorem \cite{Crooks1999} for quantum bipartite systems, which is a microscopic description of Eq. \eqref{main}. Furthermore, by definition of time-reversibility, a time-reversed evolution of an isolated quantum system $\rho_{AB}$ must be $U^{\dagger}_{AB} \tilde{\rho}_{AB}^{\rm i} U_{AB} = \rho_{AB}^{\rm i}$, where $U=U_{AB} \otimes I_R$. This constraint assigns the initial state of the time-reversed process as $\tilde{\rho}_{AB}^{\rm i} = \rho_{AB}^{\rm f}$ \cite{Sagawa2012aa,Esposito:2009aa}, which leads to the removal of the tilde over the letters for $\rho_{AB}$.

%\footnote{In general, due to $\tilde{U}  = \Theta U^{\dagger} \Theta^{\dagger}$ and $\Theta\Theta^{\dagger}=\Theta^{\dagger}\Theta=1$, the time-reversed joint probability $\tilde{p}_{m',a',b',m,a,b;r',r} :=|\bra{\tilde{m}}\tilde{U}\ket{\tilde{m}'}|^2|\langle \tilde{m}| \tilde{a},\tilde{b} \rangle|^2|\langle\tilde{m}' |\tilde{a}',\tilde{b}' \rangle|^2 \tilde{p}_{m'}$ is related to the time-forward process by the fact that $|\bra{m}\Theta^{\dagger}\Theta U^{\dagger}\Theta^{\dagger}\Theta\ket{m'}|^2 |\langle m|\Theta^{\dagger}\Theta| a,b \rangle|^2 |\bra{m'}\Theta^{\dagger}\Theta\ket{a',b'}|^2 = |\bra{m}U^{\dagger}\ket{m'}|^2 |\langle m| a,b \rangle|^2 |\langle m'| a',b' \rangle|^2$.}

A caution is needed here. We must keep in mind that the time-reversed probabilities considered here are only those with indices $\{m',m,;r',r\}$ for which time-forward transitions are nonzero. In general, there can be the time-reversed transitions between the states that are not involved in the forward transitions. This is related to the notion called absolute irreversibility \cite{Funo:2015aa,Murashita:2017aa}. In general, the absolute irreversibility occurs if the support $\mathcal{H}_{\rm i}$ of the initial state does not span the whole Hilbert space $\mathcal{H}_{\rm I}$ \cite{Funo:2015aa} (See Fig. 1). In this case, the time-reversed transitions can be divided into two groups, those that return to $\mathcal{H}_{\rm i}$ and others that do not. This makes the identity $1=\sum_{\{m,r\}\in \mathcal{H}_{\rm i}} \tilde{p}_{m',a',b',m,a,b;r',r}+\sum_{\{m,r\}\notin \mathcal{H}_{\rm i}} \tilde{p}_{m',a',b',m,a,b;r',r}$ \cite{Funo:2015aa}
and those probabilities falling outside $\mathcal{H}_{\rm i}$ should be excluded to make a valid FT and its accompanying inequalities. 
Thus, the total sum of $\tilde{p}_{m',a',b',m,a,b;r',r}$ relevant to our formulation can be less than (absolute irreversibility), or equal to (no absolute irreversibility), unity, i.e., $ \gamma_{\tilde R}\equiv\sum_{\{m,r\}\in \mathcal{H}_{\rm i}} \tilde{p}_{m',a',b',m,a,b;r',r} \le1$.  This becomes apparent when we prove Eq. (5). More details are also illustrated by an example later.

\section{FT with quantum correlation}

We first introduce an integral fluctuation theorem that was originally presented by Deffner and Lutz in \cite{Deffner:2011aa}. There the authors established the FT for an open quantum system  by taking TMP in the eigenstate basis of the initial and the final density operators (See also Sec. III A) . We rederive it here by explicitly addressing quantum correlation and generalize the FT to the situation with absolute irreversibility. For an arbitrary initial bipartite state, we have
\begin{equation}
\langle e^{- \Delta s_A - \Delta s_B + \Delta I + \beta Q}\rangle = \gamma_{\tilde R}, \label{main}
\end{equation}
with $ \gamma_{\tilde R}\equiv\sum_{\{m,r\}\in \mathcal{H}_{\rm i}} \tilde{p}_{m',a',b',m,a,b;r',r} \le1$. 

Here the change in the stochastic entropy of the subsystems is $\Delta s_{A(B)} : = -\ln p_{a'(b')} - (-\ln p_{a(b)})$, the heat transferred from reservoir to system $\beta Q := \ln \tilde{p}_{r'} - \ln p_{r} = \beta[E_{r}^R - E_{r'}^R]$ with $p_r = e^{-\beta E_r^R}/Z_{\beta}$ and $\tilde{p}_{r'} =e^{-\beta E_{r'}^R}/Z_{\beta}$, and the change of the information content $\Delta I := I_{\rm f} -I_{\rm i}$. Its proof will be given in the subsection III A.  If all nonzero transitions in the time-forward process are identical  with those in the time-reversed process, the factor $ \gamma_{\tilde R}\equiv\sum_{\{m,r\}\in \mathcal{H}_{\rm i}} \tilde{p}_{m',a',b',m,a,b;r',r}$ becomes unity, i.e. $\langle e^{- \Delta s_A - \Delta s_B + \Delta I + \beta Q}\rangle =1$, which is the case treated in \cite{Deffner:2011aa}. However, this is generally not the case as we will show with an example later.

We may also address Eq. \eqref{main} in a different context. Using the technique of Ref. \cite{Quan2008,Piechocinska2000,Crooks:2008aa}, one may describe heat transferred to each subsystem $A$ and $B$ denoted as $Q_A$ and $Q_B$, and then find an additional heat in heat transfer $Q$ with $Q' := Q - (Q_A + Q_B)$, which is driven by the interaction between the subsystems.
Equation (5) is thus rewritten as
\begin{equation}
\langle e^{- \sigma_A - \sigma_B + \Delta \Gamma}\rangle = \gamma_{\tilde R}\label{main2}
\end{equation}
with $\sigma_{A(B)} := \Delta S_{A(B)} - \beta Q_{A(B)}$ (local entropy production) and $\Delta \Gamma : = \Delta I - \beta Q'$ (additional entropy production due to quantum correlation). In Eq. \eqref{main2}, we obtain $\Delta \Gamma = 0$ if neither initial correlations nor intermediate interactions between subsystems during the evolution exist. 
Thus, Eq. \eqref{main2} may be seen as an expression explicitly addressing two subsystems out of quantum FTs for a single system, i.e., $\langle e^{- \sigma}\rangle = 1$ \cite{Tasaki2000,Kurchan2000,Manzano2015}.

In the FT of Eq. (5), one may embody measurement-induced disturbances of the joint system for an unknown joint state. For example, the initial state is prepared after measurement or the final state after a fast quenching process. In order to bypass the disturbances, one might consider evasive tactics. Statistically estimating the measurement without performing it, the joint probabilities can be evaluated by the technique exploited in Ref. \cite{Allahverdyan2014}. Furthermore, either direct energy measurements on the environment with a system intact \cite{Crooks2008} or introducing an external agent such as an energy bath can be exploited based upon the quantum channel method \cite{Aberg:2018aa, Alhambra2016}.

Using Jensen's inequality from Eq. \eqref{main}, we immediately obtain 
 \begin{equation}
 \beta \langle Q \rangle \leq  \langle \Delta s_A \rangle + \langle \Delta s_B \rangle - \langle \Delta I \rangle + \ln \gamma_{\tilde R}\label{ineq}
 \end{equation}
 for the bound of heat involving quantum correlation. The statistical averages of the stochastic entropy change in the subsystems and the change in quantum mutual information content are given by $\langle \Delta s \rangle = S(\rho_{\rm f}) - S(\rho_{\rm i})$ and $\langle \Delta I \rangle = S(A:B)_{\rm f} - S(A:B)_{\rm i}$, where $S(A:B) = S(\rho_A) + S(\rho_B) - S(\rho_{AB})$ is the quantum mutual information \cite{Nielsen2000}. Inequality \eqref{ineq} describing the total heat transferred from a reservoir is bounded by the change in the local entropies of individual systems and their correlation together with the degree of absolute irreversibility quantified by $\gamma_{\tilde R}$.  % In addition, the inequality is saturated only when the ideal reversibility condition for the joint system is satisfied, i.e., $p_{m,m';r,r'} = \tilde{p}_{m',m;r',r}$ for all defined $m,m',r,r'$, which can be realized in infinite time scale \cite{BP00}. Note that it would be a violation of second law if heat exceeded the upper bound of inequality \eqref{ineq}. The inequality \eqref{ineq} thus explains the fundamental limit for heat transfer from a thermal reservoir to a quantum bipartite system. 

\subsection{Proof of FT in Eq. (5)} 
We here prove Eq. \eqref{main}. Given the joint probability for the forward process and the backward process, the detailed fluctuation theorem is shown by noting 
\begin{eqnarray}
& &p_{m,a,b,m',a',b';r,r'}\nonumber\\ 
&=& |\bra{m',r'} U \ket{m,r}|^2|\langle m | a, b \rangle|^2 |\langle m' | a', b' \rangle|^2p_{m}p_{r}\nonumber\\
&=& |\bra{m,r} U^{\dagger} \ket{m',r'}|^2 |\langle m | a, b \rangle|^2 |\langle m' | a', b' \rangle|^2\tilde{p}_{m'}\tilde{p}_{r'}\frac{p_{m}p_{r}}{\tilde{p}_{m'}\tilde{p}_{r'}}\nonumber\\
&=& \tilde{p}_{m',a',b',m,a,b;r',r}\frac{p_{a}}{\tilde{p}_{a'}}\frac{p_{b}}{\tilde{p}_{b'}}\frac{p_{m}}{p_{a}p_{b}}\frac{\tilde{p}_{a'}\tilde{p}_{b'}}{\tilde{p}_{m'}} \frac{p_{r}}{\tilde{p}_{r'}}\nonumber\\
&=& \tilde{p}_{m',a',b',m,a,b;r',r}e^{\Delta s_A}e^{\Delta s_B}e^{- \Delta I }e^{-\beta Q},\nonumber\\
\end{eqnarray}
where the third line follows from $|\bra{m',r'} U \ket{m,r}|^2 = |\bra{m,r} U^{\dagger} \ket{m',r'}|^2$ \cite{Sagawa2012aa} and the fifth line follows from the definition of $\Delta s_A :=\ln \frac{p_{a}}{\tilde{p}_{a'}},  \Delta s_B := \ln \frac{p_{b}}{\tilde{p}_{b'}}, \Delta I := -\ln \frac{\tilde{p}_{a'}\tilde{p}_{b'}}{\tilde{p}_{m'}}+ \ln \frac{p_{a}p_{b}}{p_{m}}$, and $\beta Q:= - \ln \frac{p_{r}}{\tilde{p}_{r'}}$. The detailed fluctuation theorem for a bipartite system is then given by
\begin{equation}
\frac{\tilde{p}_{m',a',b',m,a,b;r',r}}{p_{m,a,b,m',a',b';r,r'}}  = e^{- \Delta s_A - \Delta s_B + \Delta I + \beta Q}.\label{crooks}
\end{equation}
Obviously, the relation in Eq. (9) is valid only for those time-forward transitions with $p_{m,a,b,m',a',b';r,r'}>0$, i.e., $\{m,r\}\in \mathcal{H}_{\rm i}$. 
In the end, the integral fluctuation theorem, Eq. \eqref{main}, is verified by the statistical average of the detailed fluctuation relation,
\begin{eqnarray}
& &  \langle e^{- \Delta s_A - \Delta s_B + \Delta I + \beta Q}\rangle  \nonumber\\
&=& \sum_{\{m,r\}\in \mathcal{H}_{\rm i}} p_{m,a,b,m',a',b';r,r'}\frac{\tilde{p}_{m',a',b',m,a,b;r',r}}{p_{m,a,b,m',a',b';r,r'}}\nonumber\\
&=& \sum_{\{m,r\}\in \mathcal{H}_{\rm i}} \tilde{p}_{m',a',b',m,a,b;r',r}\equiv\gamma_{\tilde R}.\nonumber\\
\end{eqnarray}
%In addition, we find that 
%\begin{eqnarray}
%& &\langle \Delta s_A + \Delta s_B - \Delta I - \beta Q \rangle \nonumber\\
%&=& \sum p_{m,a,b,m',a',b';r,r'}\ln \frac{p_{m,a,b,m',a',b';r,r'}}{\tilde{p}_{m',a',b',m,a,b;r',r}}.\label{relative}\nonumber\\
%\end{eqnarray}
%The inequality \eqref{ineq} is also verified by Eq. \eqref{relative} as a consequence of the positivity of the relative entropy.

Before moving on, we mention that our approach is identical to the one used by Deffner and Lutz in \cite{Deffner:2011aa} if we cancel out the local terms $p_a,p_b,{\tilde p}_{a'},{\tilde p}_{b'}$ in Eq. (8). In that case, we can sum over the local indices to obtain the detailed relation as $\frac{\tilde{p}_{m',m;r',r}}{p_{m,m',;r,r'}}  = e^{- \Delta s_{AB}+ \beta Q}$, by which Ref. \cite{Deffner:2011aa} obtained the FT, $\langle e^{- \Delta s_{AB}+ \beta Q}\rangle=1$. In this respect, Eq. (5) rephrases their FT by explicitly addressing correlation with a generalization to include absolute irreversibility. 

\begin{figure}[ht]
\includegraphics[width=0.5\textwidth]{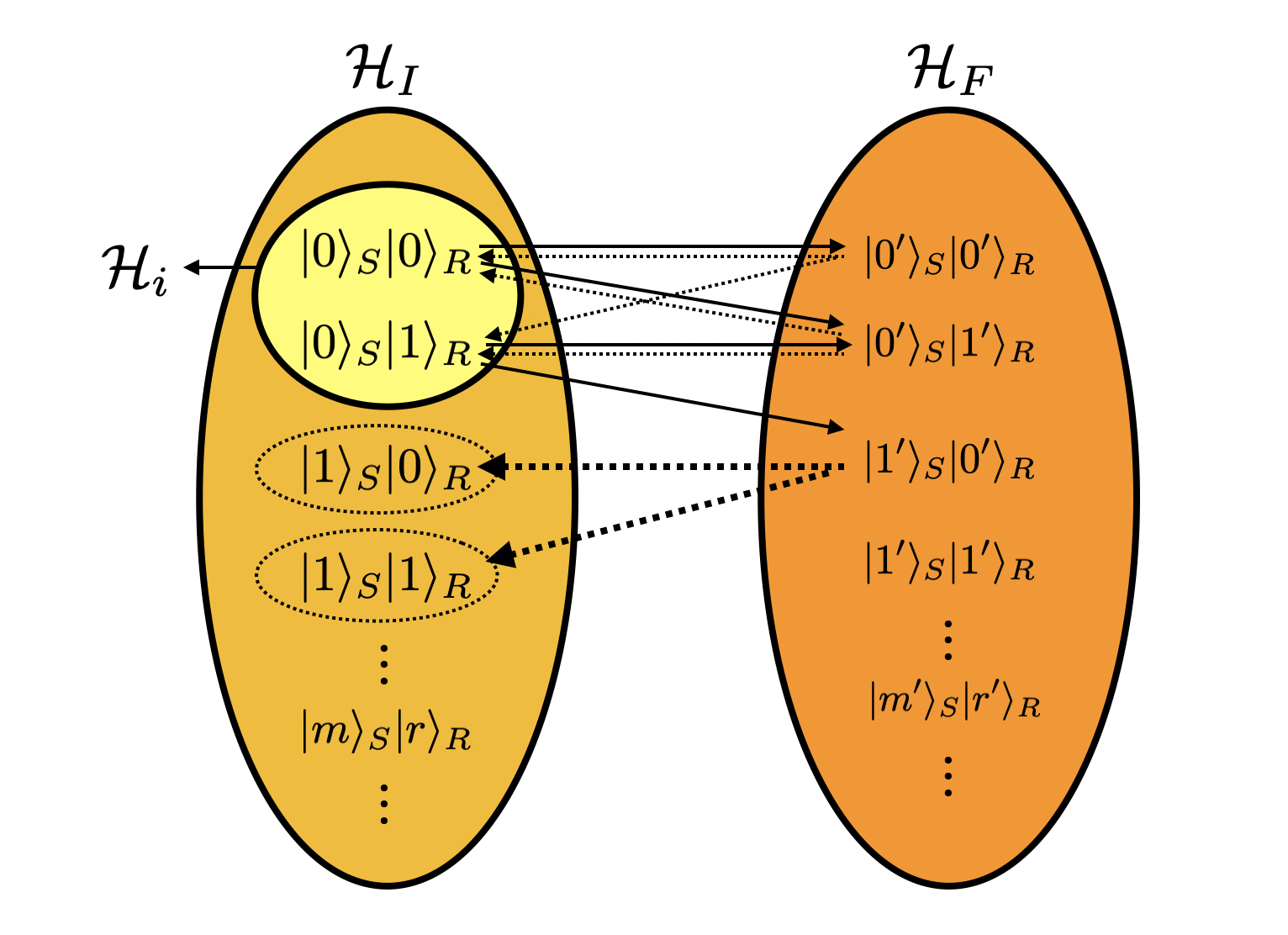}
\caption{Thermodynamic evolution in which the absolute irreversibility occurs. 
$\mathcal{H}_{\rm I}$ and $\mathcal{H}_{\rm F}$ are the whole Hilbert spaces at the initial and the final time of evolution, respectively. 
$\mathcal{H}_{\rm i}$ (light yellow) is the sub Hilbert space corresponding to the initial state, which does not span the entire space of $\mathcal{H}_{\rm I}$. 
Solid (dotted) lines represent the time-forward (reversed) transitions. The thick dotted lines represent the time-reversed transition in which the final states are outside $\mathcal{H}_{\rm i}$. }
\label{Fig:AI}
\end{figure}

\subsection{FT with quantum correlation over time-reversed process}
%When a joint system experiences dissipation, i.e., the entropy production of the joint system occurs $\langle \sigma_{tot} \rangle > 0$, where $\sigma_{tot} := \ln p_{m,m';r,r'} - \ln \tilde{p}_{m',m;r',r}$, the inequality \eqref{ineq} is no longer saturated.
%However unlike this situation, 
We now present our main results. Our approach with Eqs. \eqref{jp3} and \eqref{jp4} can provide a thermodynamic bound that is unconventionally saturated in a dissipative process. We here derive a different form of FT involving quantum correlation over the time-reversed process. 

We first use a relation, $\tilde{p}_{m',a',b',m,a,b;r',r}e^{-\Delta I} = p_{m,a,b,m',a',b';r,r'} e^{- \Delta s_A - \Delta s_B + \beta Q}$ from the detailed fluctuation relation, Eq. \eqref{crooks}. After summing the both sides over all indices, we have our main result
\begin{equation}
\langle e^{- \Delta s_A - \Delta s_B  + \beta Q}\rangle = \langle e^{-\Delta I} \rangle_{\tilde{R}}, \label{eqsub}
\end{equation}
where $\langle ... \rangle_{\tilde{R}}$ indicates an average over the time-reversed joint probability, $\tilde{p}_{m',a',b',m,a,b;r',r}$. It is worth noting that the time-reversed average of the informational component is a statistical technique firstly introduced in Ref. \cite{Sagawa2010} to show the role of information in the Jarzynski's equality using feedback control. 
The FT in Eq. (11) relates the local entropy production through the time-forward process to the change of quantum correlation through the time-reversed process.

Eq. \eqref{eqsub} also leads to a thermodynamic inequality
 \begin{equation}
 \beta \langle Q \rangle \leq  \langle \Delta s_A \rangle + \langle \Delta s_B \rangle + \ln\langle e^{-\Delta I} \rangle_{\tilde{R}}, \label{ineq2}
 \end{equation}
which is saturated if $\Delta s_A + \Delta s_B -\beta Q$ does not fluctuate. 
If the conventional second-law for a single system, i.e. $\beta \langle Q \rangle \leq  \langle \Delta s \rangle$, is extended to a bipartite system using $s_{AB}=s_A+s_B-I$, one obtains an inequality 
 \begin{equation}
 \beta \langle Q \rangle \leq  \langle \Delta s_A \rangle + \langle \Delta s_B \rangle -\langle \Delta I \rangle, \label{ineq22}
 \end{equation}
which is the inequality (7) without absolute irreversibility, $\gamma_{\tilde R}=1$.
In the saturation condition, it can be shown that  the tight bound of Eq. \eqref{ineq2} is lower than that of Eq. \eqref{ineq22}, i.e.,  $\ln\langle e^{-\Delta I} \rangle_{\tilde{R}} \leq - \langle \Delta I \rangle$, which indicates that  the statistics in the time-reversed process leads us to obtain a tighter thermodynamic bound and generalize the informational content $- \langle \Delta I \rangle$ to $\ln\langle e^{-\Delta I} \rangle_{\tilde{R}}$ in a dissipative non-equilibrium process. 

We now illustrate Eq. \eqref{ineq2} and its tightness with an example of an isothermal work extraction on each subsystem in an initially quantum correlated state. In particular, we show that our inequality \eqref{ineq2} is stronger than the conventional one in Eq. (7) for a broad set of Werner states. In feedback control processes, such non-trivial reversibility conditions have also been discussed \cite{Horowitz:2011aa}. 

\subsection{Example}

---{\it Isothermal process}: We first consider an isothermal process for a single qubit system and then apply it to each subsystem of a bipartite system both in a time-forward and in a time-backward manner. The isothermal process is performed by varying $\Delta$, the energy gap between the ground state $\ket{0}$ and the excited state $\ket{1}$, from 0 to $\Delta$ ($>> \beta^{-1}$) (Fig. 2). 
Let us here take a theoretical, though unphysical, limit $\beta\Delta\rightarrow\infty$ at the final time of transition. 
We also assume that $\Delta$ changes very slowly so that the system remains in a thermal equilibrium state at each time. 
The system then evolves from $\rho^{\rm i}_{A(B)} = \frac{1}{2}\ket{0}\bra{0}+\frac{1}{2}\ket{1}\bra{1}$ to $\rho^{\rm f}_{A(B)} = \ket{0'}\bra{0'}$, with the occupation probabilities given by $p_0^{A(B)} = \frac{1}{1 + e^{-\beta \Delta}} $ and $p_1^{A(B)} = \frac{ e^{-\beta \Delta}}{1 + e^{-\beta \Delta}}$ for each $\Delta$. (See chapter 3.B in \cite{Alicki:2004aa} for details.) 

\begin{figure}[ht]
\includegraphics[width=0.5\textwidth]{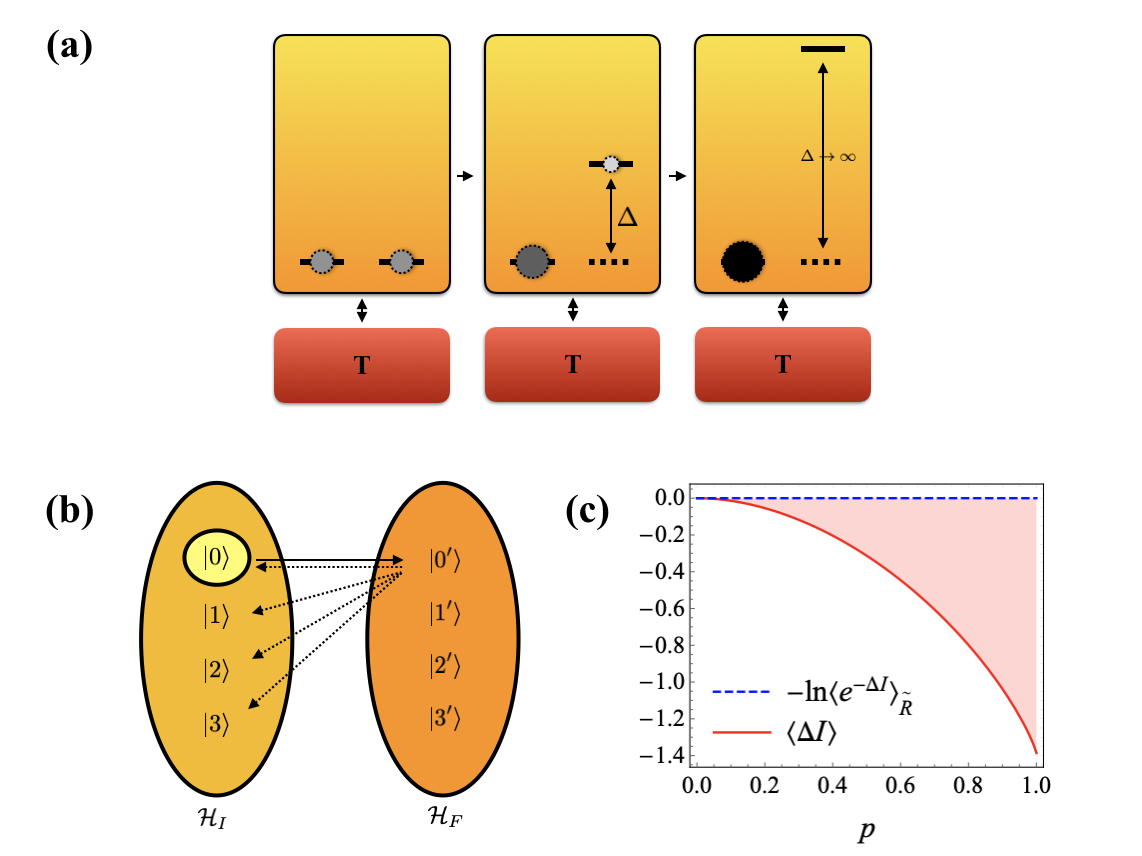}
\caption{(a) Energy-level change of each qubit under an isothermal process (b) For an initially pure Bell state $|0\rangle$ (light yellow), the absolute irreversibility occurs with $\gamma_{\tilde R}=1/4$ under the isothermal process in (a). On the other hand, for an initially mixed Werner state with $p<1$, which has all four Bell states as components,
no absolute irreversibility occurs with $\gamma_{\tilde R}=1$. (c) Difference (shaded region) between our bound $-\ln\langle e^{-\Delta I} \rangle_{\tilde{R}}=0$ in Eq. (22) and the conventional bound $\langle\Delta I \rangle$ in Eq. (21) as a function of $p$ (fraction of the pure Bell state in the initial state). The difference increases with $p$.
 }
\label{Fig:AI}
\end{figure}

---{\it Initial state}: We now apply this isothermal process to each subsystem of a Werner state, which is a mixture of Bell states \cite{Nation2012}. We use the notations for Bell states as 
\begin{eqnarray}
\ket{0}&=&\frac{1}{\sqrt2} \left(\ket{0}_A\otimes\ket{0}_B + \ket{1}_A\otimes\ket{1}_B\right) \nonumber\\
\ket{1}&=&\frac{1}{\sqrt2} \left(\ket{0}_A\otimes\ket{0}_B - \ket{1}_A\otimes\ket{1}_B\right)\nonumber\\ 
\ket{2}&=&\frac{1}{\sqrt2} \left(\ket{0}_A\otimes\ket{1}_B + \ket{1}_A\otimes\ket{0}_B\right) \nonumber\\ 
\ket{3}&=&\frac{1}{\sqrt2} \left(\ket{0}_A\otimes\ket{1}_B - \ket{1}_A\otimes\ket{0}_B\right). \nonumber\\
\end{eqnarray}
The joint system is initially prepared as $\rho_{AB}^{\rm i} =p \ket{0}\bra{0}+\frac{1-p}{4}I$, where $p$ is the fraction of Bell state $\ket{0}\bra{0}$ and $I$ the identity operator. This state can also be expressed as $\rho_{AB}^{\rm i} = \sum_{m=0}^3 p_m\ket{m}\bra{m}$ with $p_0=\frac{1+3p}{4}$ and $p_{1,2,3}=\frac{1-p}{4}$.

---{\it Final state}: At $t_{\rm i}=0$, this bipartite system, after being prepared in a Werner state, is brought to interact with a thermal reservoir at an inverse temperature $\beta$ (Fig. 2). This interaction will decohere the system so that the initial quantum correlation disappears. 
With the adiabatic change $\Delta$ of Hamiltonian, the final state of the bipartite system becomes a product state with no correlation at all. That is, $\rho_{AB}^{\rm f} =\ket{0'}\bra{0'} = \ket{0'}\bra{0'}_{A}\otimes\ket{0'}\bra{0'}_{B}$ in the product-state basis 
\begin{eqnarray} 
\ket{0'} &=& \ket{0'}_A\otimes\ket{0'}_B, \nonumber\\
\ket{1'} &=& \ket{0'}_A\otimes\ket{1'}_B, \nonumber\\
\ket{2'} &=& \ket{1'}_A\otimes\ket{0'}_B, \nonumber\\
\ket{3'} &=& \ket{1'}_A\otimes\ket{1'}_B. \nonumber\\
\end{eqnarray}

---{\it Time-reversed process}: Next, we isothermally change the gap from $\Delta=\infty$ to $0$ in a time-reversed transition. The initial and the final density operators are then given by $\tilde{\rho}_{AB}^{\rm i} = \ket{0'}\bra{0'}=\ket{0'}\bra{0'}_{A}\otimes\ket{0'}\bra{0'}_{B}$ and $\tilde{\rho}_{AB}^{\rm f} = \frac{1}{2}\left(\ket{0}\bra{0}+\ket{1}\bra{1}\right)_A \otimes \frac{1}{2}\left(\ket{0}\bra{0}+\ket{1}\bra{1}\right)_B$, respectively. The final state in the time-backward process is a completely mixed state with no correlation between subsystems $A$ and $B$, regardless of the initial state $\rho_{AB}^{\rm i} =p \ket{0}\bra{0}+\frac{1-p}{4}I$ in the time-forward process. In fact, there cannot exist correlation at any moment during the backward transition.

---{\it Transition probabilities}: In the above process, the joint probabilities in Eqs. (1), (2), (3), and (4), and the mutual information are given as follows. In Eq. (1), the joint probability $p_{m_t,a_t,b_t}= |\langle m_t | a_t, b_t \rangle|^2 p_{m_t}$ at $t_{\rm i}$ and $t_{\rm f}$ are given by 
\begin{eqnarray}\label{jpcond}
p_{0,0,0} &&= p_{0,1,1} = \frac{1}{2}\times\frac{1+3p}{4},\nonumber\\  
p_{1,0,0} &&= p_{1,1,1} = \frac{1}{2}\times\frac{1-p}{4},\nonumber\\  
p_{2,0,1} &&= p_{2,1,0} = \frac{1}{2}\times\frac{1-p}{4},\nonumber\\  
p_{3,0,1} &&= p_{3,1,0} = \frac{1}{2}\times\frac{1-p}{4}
\end{eqnarray}
and $\  p_{0',0',0'} = 1$. Otherwise, $p_{m,a,b} = p_{m',a',b'} = 0$ since the conditional probabilities $|\langle 0 | 0, 1 \rangle|^2 = |\langle 0 | 1, 0 \rangle|^2 =0$ for $m=0$, {\it etc}. Similarly, we have $|\langle 0' | 0', 1' \rangle|^2 = |\langle 0' | 1', 0' \rangle|^2 = |\langle 0' | 1', 1' \rangle|^2 =0$ for $m'=0'$ and $p_{m'} = 0$ for $m' = 1',2',3'$ at $t_{\rm f}$. 

The initial local states are both completely mixed states, $\rho_A^{\rm i}=\rho_B^{\rm i}=\frac{1}{2}I$, regardless of $p$ (fraction of the Bell state $|0\rangle$). Thus, $p_{a=0,1}=p_{b=0,1}=\frac{1}{2}$. Then the mutual information content is given by $I_{m=0,a=0,b=0} = -\ln\frac{p_{a=0}p_{b=0}}{p_{m=0}}=-\ln\frac{\frac{1}{2}\times\frac{1}{2}}{\frac{1+3p}{4}}=\ln(1+3p)$. Similarly, we obtain $I_{0,1,1}=\ln(1+3p)$ and $I_{1,0,0}=I_{1,1,1}=I_{2,0,1}=I_{2,1,0}=I_{3,0,1}=I_{3,1,0}=\ln(1-p)$.  In addition, $I_{m',a',b'} := 0$ for all $m',a',b'$ since $\rho_{AB}^{\rm f}$ is a product state. We can simply set $\ln p_{m_t} := 0$ and $I_{m_t,a_t,b_t} := 0$ for $p_{m_t}=0$.

%As a result, $\langle I_i \rangle = \sum p_{m,a,b} I_{m,a,b} = 2\ln 2$.

According to Eq. \eqref{jpcond}, the joint probability in the time forward process is given by 
\begin{eqnarray}
p_{0,0,0,0',0',0'} &&= p_{0,1,1,0',0',0'}= \frac{1+3p}{8},\nonumber\\  
p_{1,0,0,0',0',0'} &&= p_{1,1,1,0',0',0'}= \frac{1-p}{8},\nonumber\\  
p_{2,0,1,0',0',0'} &&= p_{2,1,0,0',0',0'}= \frac{1-p}{8},\nonumber\\  
p_{3,0,1,0',0',0'} &&= p_{3,1,0,0',0',0'}= \frac{1-p}{8},\nonumber 
\end{eqnarray}
and $p_{m,a,b,m',a',b'} = 0$ otherwise. Our quasi-static process yields $p_{m,m';r,r'} = p_{m,m'}p_{r,r'}$ since the system is assumed to be in equilibrium with the bath during the process. Thus, the indices of the bath $r$ and $r'$ do not affect our results.

Considering Eq. (4) and the final state in the time-reversed process, we find $\tilde{p}_{0',0',0',0,0,0} =\tilde{p}_{0',0',0',0,1,1}= \tilde{p}_{0',0',0',1,0,0} =\tilde{p}_{0',0',0',1,1,1} = \tilde{p}_{0',0',0',2,0,1}  = \tilde{p}_{0',0',0',2,1,0}= \tilde{p}_{0',0',0',3,0,1}  = \tilde{p}_{0',0',0',3,1,0} = \frac{1}{8}$, and zero otherwise.\\

---{\it Absolute irreversibility}: We now compare the thermodynamic inequalities (7) and (12) using the example in the above dissipative process.  %Bearing Eq. (9) in mind, i.e., $\beta \langle Q \rangle \leq  \langle \Delta s_A \rangle + \langle \Delta s_B \rangle + \ln\langle e^{-\Delta I} \rangle_{\tilde{R}}$, we immediately examine the validity of the equation. To end this, 
First, let us see if the absolute irreversibility occurs in this isothermal process. If $p$=1, i.e. the initial state is a pure Bell state $|0\rangle$, we have $\gamma_{\tilde R}=\sum \tilde{p}_{m',a',b',m,a,b;r',r}=\tilde{p}_{0',0',0',0,0,0}+\tilde{p}_{0',0',0',0,1,1}=\frac{1}{4}<1$. Once again, the sum over the time-reversed transition probabilities must involve only those terms with indices $\{m',m,;r',r\}$ for which time-forward transitions are nonzero. Therefore only the configurations $\{m',a',b',m,a,b\}=\{0,0,0,0,0,0\}$ and $\{0,0,0,0,1,1\}$ are included in the sum, representing the absolute irreversibility. In contrast, for all other Werner states $\rho_{AB}^{\rm i} =p \ket{0}\bra{0}+\frac{1-p}{4}I$ with $p<1$, there is no absolute irreversibility because the initial state has all components of $m=0,1,2,3$ states, thus $\gamma_{\tilde R}=1$ (Fig. 2 (b).

---{\it Information through reverse transition}: Let us now compute the informational content $\ln\langle e^{-\Delta I} \rangle_R$ during the time-reversed process. For an initial pure Bell state with $p=1$, we have
\begin{eqnarray}
&&\langle e^{-\Delta I} \rangle_{\tilde{R}}\nonumber\\
&=& \sum_{m',a',b',m,a,b} \tilde{p}_{m',a',b',m,a,b} e^{-\Delta I} \nonumber\\
&=& \tilde{p}_{0',0',0',0,0,0} e^{I_{0,0,0} - I_{0',0',0'}}+\tilde{p}_{0',0',0',0,1,1} e^{I_{0,1,1} - I_{0',0',0'}}\nonumber\\
&=& \frac{1}{8} \times e^{2\ln2} +\frac{1}{8} \times e^{2\ln2}  = 1.\nonumber\\ 
\end{eqnarray}
We thus have $\ln\langle e^{-\Delta I} \rangle_{\tilde{R}} = 0$. On the other hand, for an initial mixed state with $p<1$, we have
\begin{eqnarray}
&&\langle e^{-\Delta I} \rangle_{\tilde{R}}\nonumber\\
=&& \sum_{m',a',b',m,a,b} \tilde{p}_{m',a',b',m,a,b} e^{-\Delta I} \nonumber\\
=&& \tilde{p}_{0',0',0',0,0,0} e^{I_{0,0,0} - I_{0',0',0'}}+\tilde{p}_{0',0',0',0,1,1} e^{I_{0,1,1} - I_{0',0',0'}}\nonumber\\
&&+\tilde{p}_{0',0',0',1,0,0} e^{I_{1,0,0} - I_{0',0',0'}}+\tilde{p}_{0',0',0',1,1,1} e^{I_{1,1,1} - I_{0',0',0'}}\nonumber\\
&&+\tilde{p}_{0',0',0',2,0,1} e^{I_{2,0,1} - I_{0',0',0'}}+\tilde{p}_{0',0',0',2,1,0} e^{I_{2,1,0} - I_{0',0',0'}}\nonumber\\
&&+\tilde{p}_{0',0',0',3,0,1} e^{I_{3,0,1} - I_{0',0',0'}}+\tilde{p}_{0',0',0',3,1,0} e^{I_{3,1,0} - I_{0',0',0'}}\nonumber\\
&=& 1. 
\end{eqnarray}
Therefore, we obtain $\ln\langle e^{-\Delta I} \rangle_R=1$ regardless of whether the initial state is a pure or mixed Werner state.
 %\textcolor{blue}{We note that, except for the configurations $\{0,0,0,0,0,0\}$ and $\{0,0,0,0,1,1\}$, $\tilde{p}_{m',a',b',m,a,b}$ are not included since they are prohibited in the time-forward transitions. The time-reversed average can be reaffirmed by the fact that $\avr{e^{-\Delta s_A - \Delta s_B +\beta Q}} =1$ which is always valid for an isothermal process applied to each subsystem.}

---{\it Heat exchange}: The heat transferred from a heat bath to a two-level system equals $Q=-\beta^{-1} \ln 2$ during an isothermal process applied to each subsystem. This can be readily seen by making an analogy to a particle in a box. During the process, the total energy of each system is the same at the initial and the final time whereas an external work is done with an amount $W=\beta^{-1} \ln 2$ (particle only on one side in the final state), thus $Q=\Delta E-W=-\beta^{-1} \ln 2$. Therefore the total amount of heat transfer is given by 
\begin{eqnarray}
 \langle Q \rangle =-2 \times \beta^{-1} \ln 2.\label{} 
\end{eqnarray}
---{\it Entropy change}: The entropy change in the subsystem $A(B)$ is given by 
\begin{eqnarray}
\langle \Delta s_{A(B)} \rangle = -\ln 2,\nonumber\label{}
\end{eqnarray}
where we use the initial and the final state given by $\rho_{A(B)}^{\rm i} = \frac{1}{2}\left(\ket{0}\bra{0}+\ket{1}\bra{1}\right)_{A(B)}$ and $\rho_{AB}^{\rm f} = \ket{0'}\bra{0'}_{A}\otimes\ket{0'}\bra{0'}_{B}$, respectively.
%In consequence, the verification of Eq. (9) with its saturation is shown by
%\begin{eqnarray}
%2 \ln 2 \leq  \ln 2  + \ln 2 + 0 .\label{}
%\end{eqnarray}

---{\it Mutual-information change}: Finally, we compute the average of quantum mutual information content $\langle \Delta I \rangle$. As the final state has no correlation at all, we only need the mutual information of the initial state. Thus, we have.
\begin{eqnarray}
 &&\langle \Delta I \rangle =\langle I_{\rm f}\rangle-\langle I_{\rm i}\rangle=0-(s_A^{\rm i}+ s_B^{\rm i}-s_{AB}^{\rm i})\nonumber\\
&&=-2\ln2-\frac{1+3p}{4}\ln\frac{1+3p}{4}-\frac{3(1-p)}{4}\ln\frac{1-p}{4}.
\end{eqnarray}

Putting all together, we are now in a position to check the inequalities (7) and (12). 
In order to show how the sum of local entropy production is bounded by the information content, we express inequalities (7) and (12) as 
\begin{eqnarray}
\langle \Delta s_A \rangle + \langle \Delta s_B \rangle-\beta \langle Q \rangle \ge \langle \Delta I \rangle - \ln \gamma_{\tilde R},\\
\langle \Delta s_A \rangle + \langle \Delta s_B \rangle -\beta \langle Q \rangle \ge -\ln\langle e^{-\Delta I} \rangle_{\tilde{R}},
\end{eqnarray}
respectively. For an initial pure Bell state with $p=1$, we obtain
\begin{eqnarray}
\langle \Delta s_A \rangle + \langle \Delta s_B \rangle-\beta \langle Q \rangle=-\ln 2  - \ln 2 + 2 \ln 2=0,\nonumber\\
\label{}
\end{eqnarray}
and 
\begin{eqnarray}
\langle\Delta I \rangle - \ln \gamma_{\tilde R}&&=-2\ln 2+2\ln2=0,\nonumber\\
-\ln\langle e^{-\Delta I} \rangle_{\tilde{R}}&&=0
\label{}
\end{eqnarray}
respectively. 
Therefore, we see that both of the inequalities are saturated even for a dissipative process. 

On the other hand, for an initial mixed Werner state with $p<1$, we obtain $\langle \Delta s_A \rangle + \langle \Delta s_B \rangle-\beta \langle Q \rangle=0$,
whereas 
\begin{eqnarray}
\langle\Delta I \rangle - \ln \gamma_{\tilde R}&&=\langle\Delta I \rangle+0,\nonumber\\
-\ln\langle e^{-\Delta I} \rangle_{\tilde{R}}&&=0
\label{}
\end{eqnarray}
respectively. Since $\langle \Delta I \rangle<0$ for any $p<1$ in Eq. (20), we see that our inequality (12) is saturated and provides a stronger bound for the local entropy production than the conventional thermodynamic inequality (7). 
In our example of Werner state, the difference between our bound $-\ln\langle e^{-\Delta I} \rangle_{\tilde{R}}$ and the conventional bound $\langle\Delta I \rangle - \ln \gamma_{\tilde R}$ increases as a function of $p$, that is, 
as a function of initial correlation (Fig. 2 (c)).

The conventional inequality (21), aside from the absolute irreversibility, can be interpreted as the mutual information lowering the bound on the local entropy production. Our example shows that there can exist a tighter bound involving initial correlation and our bound $-\ln\langle e^{-\Delta I} \rangle_{\tilde{R}}$ provides such a tighter bound in this example.

\subsection{Remarks} 
---{\it Conventional TMP:} We discuss the fluctuation theorem derived by the conventional TMP, which can be regarded as a special case. Assuming the eigenstates of the joint system are the product states of the eigenstates of the subsystems at the initial and the final time, i.e., $\ket{m}=\ket{a''}\otimes\ket{b''}=\ket{a'',b''}$ and $\ket{m'}=\ket{a'''}\otimes\ket{b'''}=\ket{a''',b'''}$, we  obtain that $p_{m,a,b,m',a',b';r,r'} = p_{a,b,a',b';r,r'}$, since $|\langle m | a, b \rangle|^2 = |\langle a'', b'' | a, b \rangle|^2 = \delta_{a,a''}\delta_{b,b''}$ and $|\langle m' | a', b' \rangle|^2 = |\langle a''',b''' | a', b' \rangle|^2 = \delta_{a',a'''}\delta_{b',b'''}$. The resulting probability $p_{a,b,a',b';r,r'}$ is equal to the probability obtained by the TMP with the measurements taken in the eigenbasis of each subsystem. As a result, Eq. \eqref{main} reduces to 
\begin{equation}
\langle e^{- \Delta s_A - \Delta s_B + \Delta J + \beta Q}\rangle = \gamma_{\tilde{R}}\label{classical},
\end{equation}
where $\Delta J = J_f - J_i$ is the change in the classical mutual information and the statistical average is performed by $p_{a,b,a',b';r,r'}$. Equation \eqref{classical} shows the fluctuation theorems for bipartite systems when the joint state does not possess quantum correlations \cite{Jevtic2015,Morikuni:2011aa,Funo2013,Manzano:2018aa}.

---{\it Heat transfer in the presence of a classical memory:} Consider now a classical observer of subsystem $B$ and a classical memory of $A$. The state of $B$ is assumed to be unchanged and no energy is allowed to be exchanged with external systems. Equation \eqref{classical} then reproduces the main result in Ref. \cite{Sagawa2012}, $\langle e^{- \sigma + \Delta J}\rangle = 1$ assuming $\gamma_{\tilde{R}}=1$,  which leads to 
\begin{equation}
\beta \langle Q \rangle \leq \langle \Delta s_A \rangle  - \langle \Delta J\rangle, \label{cLandauer}
\end{equation} where  $\sigma:=\Delta s_A -\beta Q$. Moreover, inequality \eqref{cLandauer} can be interpreted as Landauer's principle in the presence of classical correlations between a memory and an observer. In our quantum mechanical setting, inequality \eqref{ineq} becomes
\begin{equation}
\beta \langle Q \rangle \leq \langle \Delta s_A \rangle  - \langle \Delta I\rangle+\ln\gamma_{\tilde{R}}, \label{qLandauer}
\end{equation} 
when the state of the observer $B$ is assumed to be degenerate and unchanged, i.e., $\langle \Delta s_B \rangle = 0$. Inequality \eqref{qLandauer} indicates the quantum mechanical bound for the heat dissipation can reach negative values during the process of erasing memory if the memory is quantum-mechanically correlated with an observer. This is further enhanced when the absolute irreversibility occurs from the negative term $\ln\gamma_{\tilde{R}}$. This implies the environment can be cooled in the course of erasure of memory in accordance with the result in Refs. \cite{Park2013,Rio:2011aa}. 
Furthermore, under the same situation, our new inequality (12) gives 
\begin{eqnarray}
 \beta \langle Q \rangle \leq  \langle \Delta s_A \rangle + \ln\langle e^{-\Delta I} \rangle_{\tilde{R}}.
\end{eqnarray}
The negativity of the heat $\beta \langle Q \rangle$ can be greater in Eq. (28) or (29) according to the comparison between $ \langle \Delta I\rangle-\ln\gamma_{\tilde{R}}$ and $- \ln\langle e^{-\Delta I} \rangle_{\tilde{R}}$.

---{\it Comparison between information terms:} In the previous subsection, we have demonstrated an example for which our information term offers a tighter bound for the local entropy production, i.e., $- \ln\langle e^{-\Delta I} \rangle_{\tilde{R}}>\langle \Delta I\rangle-\ln\gamma_{\tilde{R}}$. This is not always the case and we here provide a counterexample. 
Let us consider unitary dynamics with a time-dependent Hamiltonian that adiabatically changes the initial product states $|0\rangle_{\rm p}\equiv|0\rangle_A|0\rangle_B$, $|1\rangle_{\rm p}\equiv|0\rangle_A|1\rangle_B$, $|2\rangle_{\rm p}\equiv|1\rangle_A|0\rangle_B$ and $|3\rangle_{\rm p}\equiv|1\rangle_A|1\rangle_B$ to the Bell states $|0\rangle,|1\rangle,|2\rangle$, and $|0\rangle$ introduced in Eq. (14). 
With an initial mixed state $\rho_{\rm i}=\frac{1+3p}{4}|0\rangle\langle0|_{\rm p}+\frac{1-p}{4}\sum_{j=1}^3|j\rangle\langle j|_{\rm p}$, a straightforward calculation gives
\begin{eqnarray}
\langle \Delta I\rangle&&=(1+p)\ln(1+p)+(1-p)\ln(1-p)\nonumber\\
\langle e^{-\Delta I} \rangle_{\tilde{R}}&&=\frac{1+p-p^2}{(1+p)^2(1-p)}\nonumber
\end{eqnarray}
with $\gamma_{\tilde{R}}=1$. It turns out that $- \ln\langle e^{-\Delta I} \rangle_{\tilde{R}}<\langle \Delta I\rangle$ for $0<p<1$.

---{\it Work statistics:} Finally, let us discuss the aspect of work statistics viewed from our approach. Numerous works addressed the role of quantum correlation in extracting work from a quantum bipartite system. In particular, Refs. \cite{Zurek:2003aa, Brodutch:2010aa} showed how the quantum discord can be exploited to extract work under one-way communication. Ref. \cite{Horodecki:2003aa} defined as a measure of quantum correlation the work deficit, i.e. the difference in extractable work between accessing the local systems with communications and accessing the global system. 
In addition, Funo {\it et al.} rigorously studied the work extraction by coupling a quantum bipartite system to a measuring device and performing the feedback dynamics according to the measurement outcome \cite{Funo:2013aa}. On the other hand, our approach does not directly deal with the feedback control, but still offers an insight into the role of quantum correlation when the quantum system undergoes a time-dependent Hamiltonian dynamics. For instance, let us address the case in which the work is extracted locally by each subsystem. 
Similar to the inequalities (7) and (12), we obtain the sum of works $W_A+W_B$ bounded as
\begin{eqnarray}
&&W_A+W_B\le -\Delta F_A-\Delta F_B-\frac{1}{\beta} \langle \Delta I\rangle+\frac{1}{\beta}\ln\gamma_{\tilde{R}},\nonumber\\
&&W_A+W_B\le -\Delta F_A-\Delta F_B + \frac{1}{\beta}\ln\langle e^{-\Delta I} \rangle_{\tilde{R}}.
\end{eqnarray}
Here $\Delta F_A $ and $\Delta F_B$ represent the change in free energy of subsystems $F_{A(B)} := U_{A(B)} - \beta^{-1}S_{A(B)}$. Thus, we see that the extraction of work can be enhanced by using the initial correlation in both of the inequalities. One may provide a tighter bound than the other according to the information terms depending on the physical situation.

\section{Summary}

In this paper, we have derived an information fluctuation theorem for a quantum bipartite system possessing an initial correlation of subsystems interacting with a thermal bath. Our fluctuation theorem holds for an arbitrary bipartite system with the use of the quantum mutual information content $I$ and the statistics obtained by the TMP. Importantly, only the global states are subject to the two-point measurements whereas the local states are incorporated in order to augment information on the correlation of the composite system without disturbing correlation via measurement. This hybrid approach enables us to identify the role of quantum correlation intact in the thermodynamic process of open quantum systems. 
We first showed how our approach can explicitly incorporate correlation by generalizing the conventional FT to the situation with absolute irreversibility.  
We then obtained a novel form of FT that relates the local entropy production in the time-forward transition to the change of correlation in the time-reversed transition.
This particularly leads to a new thermodynamic bound for heat transferred from a reservoir in the dissipative and the non-dissipative regimes. We illustrated the usefulness of our approach by an example of Werner states under an isothermal process for which we demonstrated that our new inequality is stronger than the conventional one. 

We hope that our thermodynamic approach could shed light on the nonclassical features in non-equilibrium quantum dynamics of multipartite systems. Specifically, we have derived a FT that takes into account the quantum correlation over the time-reversed process explicitly. This approach seems promising to investigate the quantum features of quantum thermodynamic processes, so will be further studied in more detail elsewhere. In this paper, we have shown there are thermodynamic processes for which our new inequality provides a tighter bound for the local entropy production while we also showed there can be other cases where the conventional one may be stronger. We thus think it worthwhile to fully investigate under which conditions one is tighter than the other. In addition, our study can be further extended to quantum multipartite systems. The role played by the so-called absolute irreversibility can be more clarified along this line in examining thermodynamic constraints of the quantum multipartite transitions.

\textit{Acknowledgments}: 
J.J.P. is supported by a KIAS Individual Grant (CG075501) at Korea Institute for Advanced Study. S.W.K. acknowledges funding from Basic Science Research Program through the National Research Foundation of Korea (NRF) (2016R1A4A1008978 and 2016R1A2B4015978). V.V. ac- knowledges funding from the John Templeton Foundation, the Eutopia Foundation, and the National Research Foundation, Prime Ministers Office, Singapore, under its Competitive Research Programme (CRP Award No. NRF- CRP14-2014- 02) and administered by Centre for Quantum Technologies, National University of Singapore.

%\bibliography{FTforarbitrarybipartite}


\end{document}